\documentclass{ws-procs10x7}
\usepackage{balance}

\newcolumntype{d}[1]{D{.}{.}{#1}}

\def\be{\begin{eqnarray}}
\def\en{\end{eqnarray}}

\def\ra{\rangle}

\makeindex
\begin{document}

\title{MIXING OF SCALAR GLUEBALL AND SCALAR QUARKONIA}

\author{HAI-YANG CHENG$^*$}

\address{Institute of Physics, Academia Sinica, Taipei, Taiwan 115, Republic of China\\
$^*$E-mail: phcheng@phys.sinica.edu.tw}


\twocolumn[\maketitle\abstract{The isosinglet scalar mesons
$f_0(1710)$, $f_0(1500)$, $f_0(1370)$ and their mixing are
studied. Two recent lattice results are employed as the starting
point; one is the approximate SU(3) symmetry in the scalar sector
above 1 GeV for the connected insertion part without $q\bar q$
annihilation, and the other is the scalar glueball mass at 1710
MeV in the quenched approximation. In the SU(3) symmetry limit,
$f_0(1500)$ becomes a pure SU(3) octet and is degenerate with
$a_0(1450)$, while $f_0(1370)$ is mainly an SU(3) singlet with a
slight mixing with the scalar glueball which is the primary
component of $f_0(1710)$. These features remain essentially
unchanged even when SU(3) breaking is taken into account.  The
observed enhancement of $\omega f_0(1710)$ production over $\phi
f_0(1710)$ in hadronic $J/\psi$ decays and the copious $f_0(1710)$
production in radiative $J/\psi$ decays lend further support to
the prominent glueball nature of $f_0(1710)$. }
 ]

\section{INTRODUCTION}

Among the isosinglet scalar mesons $f_0(1710)$, $f_0(1500)$ and
$f_0(1370)$, it has been quite controversial as to which of these
is the dominant scalar glueball. It has been suggested that
$f_0(1500)$ is primarily a scalar glueball~\cite{Close1}, due
partly to the fact that $f_0(1500)$, discovered in $p\bar{p}$
annihilation at LEAR, has decays to $\eta\eta$ and $\eta\eta'$
which are relatively large compared to that of
$\pi\pi$~\cite{ams95} and that the earlier quenched lattice
calculations~\cite{bsh93} predict the scalar glueball mass to be
around $1550$ MeV. Furthermore, because of the small production of
$\pi\pi$ in $f_0(1710)$ decay compared to that of $K\bar K$, it
has been thought that $f_0(1710)$ is primarily $s\bar s$
dominated. In contrast, the smaller production rate of $K\bar K$
relative to $\pi\pi$ in $f_0(1370)$ decay leads to the conjecture
that $f_0(1370)$ is governed by the non-strange light quark
content.

Based on the above observations, a flavor-mixing scheme is
proposed \cite{Close1} to consider the glueball and $q\bar q$
mixing in the neutral scalar mesons $f_0(1710)$, $f_0(1500)$ and
$f_0(1370)$.  Best $\chi^2$ fits to the measured scalar meson
masses and their branching ratios of strong decays have been
performed in several references by Amsler, Close and
Kirk~\cite{Close1}, Close and Zhao~\cite{Close2}, and He {\it et
al.}~\cite{He}. A typical mixing matrix in this scheme is
\cite{Close2}
 \begin{eqnarray} \label{eq:Close}
 \left(\begin{matrix} f_0(1370) \cr f_0(1500) \cr f_0(1710) \cr\end{matrix}\right)=
\left( \begin{matrix} -0.91 & -0.07 & 0.40 \cr
                 -0.41 & 0.35 & -0.84 \cr
                0.09 & 0.93 & 0.36 \cr
                  \end{matrix}\right)\left(\begin{matrix}|N\rangle \cr
 |S\rangle \cr |G\rangle \cr\end{matrix}\right). \nonumber
 \end{eqnarray}
A common feature of these analyses is that, before mixing, the
$s\bar{s}$ mass $M_S$ is larger than the glueball mass $M_G$
which, in turn, is larger than the
$N(\equiv(u\bar{u}+d\bar{d})/\sqrt{2})$ mass $M_N$, with $M_G$
close to 1500 MeV and $M_S-M_N$ of the order of $200\sim 300$ MeV.
However, there are at least four serious problems with this
scenario: (i) The isovector scalar meson $a_0(1450)$ is now
confirmed to be the $q\bar{q}$ meson in the lattice
calculation~\cite{Mathur}. As such, the degeneracy of $a_0(1450)$
and $K_0^*(1430)$, which has a strange quark, cannot be explained
if $M_S$ is larger than $M_N$ by $\sim 250$ MeV. (ii) The most
recent quenched lattice calculation with improved action and
lattice spacings extrapolated to the continuum favors a larger
scalar glueball mass close to 1700 MeV~\cite{Chen,MP}. (iii) If
$f_0(1710)$ is dominated by the $s\bar s$ content, the decay
$J/\psi\to \phi f_0(1710)$ is expected to have a rate larger than
that of $J/\psi\to \omega f_0(1710)$. Experimentally, it is other
way around: the rate for $\omega f_0(1710)$ production is about 6
times that of $J/\psi\to \phi f_0(1710)$. (iv) It is well known
that the radiative decay $J/\psi\to \gamma f_0$ is an ideal place
to test the glueball content of $f_0$. If $f_0(1500)$ has the
largest scalar glueball component, one expects  the
$\Gamma(J/\psi\to \gamma f_0(1500))$ decay rate to be
substantially larger than that of $\Gamma(J/\psi\to \gamma
f_0(1710))$. Again, experimentally, the opposite is true. Simply
based on the above experimental observations, one will naively
expect that $\gamma\gg\alpha>\beta$ in the wave function of
$|f_0(1710)\ra=\alpha|N\ra+\beta|S\ra+\gamma|G\ra$.

In our recent work \cite{CCL}, we have employed two recent lattice
results as the input for the mass matrix which is essentially the
starting point for the mixing model between scalar mesons and the
glueball. First of all, an improved quenched lattice calculation
of the glueball spectrum at the infinite volume and continuum
limits based on much larger and finer lattices have been carried
out~\cite{Chen}. The mass of the scalar glueball is calculated to
be $m(0^{++})=1710\pm50\pm 80$ MeV.   This suggests that $M_G$
should be close to 1700 MeV rather than 1500 MeV from the earlier
lattice calculations~\cite{bsh93}. Second, the recent quenched
lattice calculation of the isovector scalar meson $a_0$ mass has
been carried out for a range of low quark masses~\cite{Mathur}. It
is found that, when the quark mass is smaller than that of the
strange, $a_0$ mass is almost independent of the quark mass, in
contrast to those of $a_1$ and other hadrons that have been
calculated on the lattice (see Fig. 1).  The chiral extrapolated
mass $a_0 = 1.42 \pm 0.13$ GeV suggests that $a_0(1450)$ is a
$q\bar{q}$ state.  Furthermore, $K_0^{*}(1430)^+$, an $u\bar{s}$
meson, is calculated to be $1.41 \pm 0.12$ GeV and the
corresponding scalar $\bar{s}s$ state from the connected insertion
is $1.46 \pm 0.05$ GeV. This explains the fact that
$K_0^{*}(1430)$ is basically degenerate with $a_0(1450)$ despite
having one strange quark. This unusual behavior is not understood
as far as we know and it serves as a challenge to the existing
hadronic models. In any case, these lattice results hint at an
SU(3) symmetry in the scalar meson sector. Indeed, the near
degeneracy of $K_0^*(1430)$, $a_0(1470)$, and $f_0(1500)$ implies
that, to first order approximation, flavor SU(3) is a good
symmetry for the scalar mesons above 1 GeV.

\begin{figure}[t]
\centerline{\psfig{file=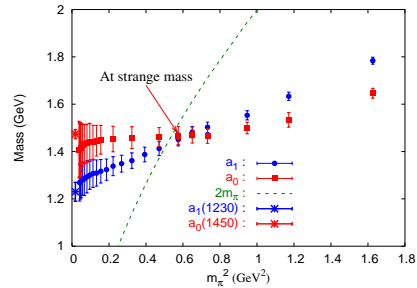,width=2.2in}} \caption{Lattice
calculations of $a_0$ and $a_1$ masses as a function of $m_\pi^2$
[6].} \label{fig1}
\end{figure}

\section{Mixing Matrix}
 We shall use $|U\ra, |D\ra, |S\ra$ to denote the
quarkonium states $|u\bar u\ra, |d\bar d\ra$ and $|s\bar s\ra$,
and $|G\ra$ to denote the pure scalar glueball state. In this
basis, the mass matrix reads
 \be \label{eq:massmatrix}
 {\rm M}=\left( \begin{matrix} M_U & 0 & 0 & 0  \cr
                   0 & M_D & 0 & 0  \cr
                    0 & 0 & M_S & 0  \cr
                      0 & 0 & 0 & M_G  \cr
                  \end{matrix} \right)+\left(\begin{matrix} x & x & x_s & y \cr
                x & x & x_s & y \cr
                x_s & x_s & x_{ss} & y_s \cr
                y & y & y_s & 0 \cr\end{matrix} \right),
                \nonumber
 \en
where the parameter $x$ denotes the mixing between different
$q\bar{q}$ states through quark-antiquark annihilation and $y$
stands for the glueball-quarkonia mixing strength. Possible SU(3)
breaking effects are characterized by the subscripts ``$s$" and
``$ss$". As noticed in passing, lattice calculations~\cite{Mathur}
of the $a_0(1450)$ and $K_0^*(1430)$ masses indicate a good SU(3)
symmetry for the scalar meson sector above 1 GeV. This means that
$M_S$ should be close to $M_U$ and $M_D$. Also the glueball mass
$M_G$ should be close to the scalar glueball mass $1710\pm50\pm80$
MeV from the lattice QCD calculation in the pure gauge sector
\cite{Chen}.

We shall begin by considering exact SU(3) symmetry as a first
approximation, namely, $M_S=M_U=M_D=M$ and $x_s=x_{ss}=x$ and
$y_s=y$. In this case, two of the mass eigenstates are identified
with $a_0(1450)$ and $f_0(1500)$ which are degenerate with the
mass $M$. Taking $M$ to be the experimental mass of $1474\pm 19$
MeV \cite{PDG}, it is a good approximation for the mass of
$f_0(1500)$ at $1507\pm 5$ MeV \cite{PDG}. Thus, in the limit of
exact SU(3) symmetry, $f_0(1500)$ is the SU(3) isosinglet octet
state $|f_{\rm octet}\ra$ and is degenerate with $a_0(1450)$. In
the absence of glueball-quarkonium mixing, i.e. $y=0$, $f_0(1370)$
becomes a pure SU(3) singlet $|f_{\rm singlet}\ra$ and $f_0(1710)$
the pure glueball $|G\ra$. The $f_0(1370)$ mass is given by
$m_{f_0(1370)}=M+3x$. Taking the experimental $f_0(1370)$ mass to
be $1370$ MeV, the quark-antiquark mixing matrix element $x$
through annihilation is found to be $-33$ MeV.  When the
glueball-quarkonium mixing $y$ is turned on, there will be some
mixing between the glueball and the SU(3)-singlet $q\bar{q}$ . If
$y$ has the same magnitude as $x$, i.e. $33$ MeV, then $3y^2 \ll
\Delta^2$ where $\Delta$ is half of the mass difference between
$M_G$ and $M+3x$, which is $\sim 170$ MeV. In this case, the mass
shift of $f_0(1370)$ and $f_0(1710)$ due to mixing is only $\sim
3y^2/2\Delta = 9.6$ MeV. In the wavefunctions of the mixed states,
the coefficient of the minor component is of order
$\sqrt{3}y/(2\Delta) = 0.17$ which corresponds to $\sim 3\%$
mixing.

 As discussed before, SU(3) symmetry leads naturally to the near
degeneracy of $a_0(1450)$, $K_0^*(1430)$ and $f_0(1500)$. However,
in order to accommodate the observed branching ratios of strong
decays, SU(3) symmetry must be broken to certain degree in the
mass matrix and/or in the decay amplitudes. One also needs
$M_S>M_U=M_D$ in order to lift the degeneracy of $a_0(1450)$ and
$f_0(1500)$.

To explain the large disparity between $\pi\pi$ and $K\bar K$
production in scalar glueball decays, Chanowitz \cite{Chanowitz}
advocated that a pure scalar glueball cannot decay into
quark-antiquark in the chiral limit, i.e.
\begin{equation}  \label{eq:chiral_supp}
 A(G\to q\bar q)\propto m_q.
 \end{equation}
Since the current strange quark mass is an order of magnitude
larger than $m_u$ and $m_d$, decay to $K\bar K$ is largely favored
over $\pi\pi$. Furthermore, it has been pointed out that chiral
suppression will manifest itself at the hadron level~\cite{Chao}.
To this end, it is suggested \cite{Chao} that $m_q$ in Eq.
(\ref{eq:chiral_supp}) should be interpreted as the scale of
chiral symmetry breaking since chiral symmetry is broken not only
by finite quark masses but is also broken spontaneously.
Consequently, chiral suppression for the ratio $\Gamma(G\to
\pi\pi)/\Gamma(G\to K\bar K)$ is not so strong as the current
quark mass ratio $m_u/m_s$.

Guided by the lattice calculations for chiral suppression in $G\to
PP$ \cite{Sexton}, the fitted masses and branching ratios are
summarized in Table \ref{tab:fit}, while the predicted decay
properties of scalar mesons are exhibited in Table
\ref{tab:prediction}. The mixing matrix obtained in our model has
the form:
 \be \label{eq:wf}
 \left(\begin{matrix} f_0(1370) \cr f_0(1500) \cr f_0(1710) \cr\end{matrix}\right)=
\left( \begin{matrix} 0.78 & 0.51 & -0.36 \cr
                 -0.54 & 0.84 & 0.03 \cr
                0.32 & 0.18 & 0.93 \cr
                  \end{matrix}\right)\left(\begin{matrix}|N\rangle \cr
 |S\rangle \cr |G\rangle \cr\end{matrix}\right). \nonumber
 \en
It is evident that $f_0(1710)$ is composed primarily of the scalar
glueball, $f_0(1500)$ is close to an SU(3) octet, and $f_0(1370)$
consists of an approximate SU(3) singlet with some glueball
component ($\sim 10\%$). Unlike $f_0(1370)$, the glueball content
of $f_0(1500)$ is very tiny because an SU(3) octet does not mix
with the scalar glueball.

\begin{table}[t]
\tbl{Fitted masses and branching ratios. The chiral suppression in
$G\to PP$ decay is taken to be $r_s=1.55$. \label{tab:fit}}
{\begin{tabular}{|c|c|c| } \hline
& \raisebox{0pt}[10pt][6pt] {Experiment} & fit   \\
\hline \raisebox{0pt}[10pt][6pt] {$M_{f_{0}(1710)}$(MeV)} &
$1718\pm6$ & 1718  \\ \hline
 $M_{f_{0}(1500)}$(MeV)&$1507\pm5$
 & 1504 \\ [3pt] \hline
\raisebox{0pt}[10pt][6pt]{$M_{f_{0}(1370)}$(MeV)}&$1350\pm 150$ &
1346  \\ [3pt] \hline
\raisebox{0pt}[10pt][6pt]{$\frac{\Gamma(f_{0}(1500)\rightarrow
\eta\eta)}{\Gamma(f_{0}(1500)\rightarrow \pi\pi)}$} &$0.145\pm0.027$  & 0.081 \\
[3pt] \hline
\raisebox{0pt}[10pt][6pt]{$\frac{\Gamma(f_{0}(1500)\rightarrow
K\bar{K})}{\Gamma(f_{0}(1500)\rightarrow \pi\pi)}$} & $0.246\pm0.026$ & 0.27 \\
[3pt] \hline
\raisebox{0pt}[10pt][6pt]{$\frac{\Gamma(f_{0}(1710)\rightarrow
\pi\pi)}{\Gamma(f_{0}(1710)\rightarrow K\bar{K})}$} & $0.30\pm0.20$  & 0.34 \\
[3pt] \hline
\raisebox{0pt}[10pt][6pt]{$\frac{\Gamma(f_{0}(1710)\rightarrow
\eta\eta)}{\Gamma(f_{0}(1710)\rightarrow K\bar{K})}$} &
$0.48\pm0.15$  & 0.51 \\
 [3pt] \hline
\raisebox{0pt}[10pt][6pt]{$\frac{\Gamma(a_{0}(1450)\rightarrow
K\bar K)}{\Gamma(a_{0}(1450)\rightarrow \pi\eta)}$} &
$0.88\pm0.23$  & 1.12 \\
[3pt]  \hline
\raisebox{0pt}[10pt][6pt]{$\frac{\Gamma(a_{2}(1320)\rightarrow
K\bar K)}{\Gamma(a_{2}(1320)\rightarrow \pi\eta)}$} &
$0.34\pm0.06$   & 0.46 \\
[3pt]  \hline
\raisebox{0pt}[10pt][6pt]{$\frac{\Gamma(f_2(1270)\rightarrow K\bar
K)}{\Gamma(f_{2}(1270)\rightarrow \pi\pi)}$} &
$0.054^{+0.005}_{-0.006}$   & 0.057 \\
[3pt]  \hline
 \raisebox{0pt}[10pt][6pt]{$\chi^2/{\rm d.o.f.}$} & & 2.5 \\
 \hline
\end{tabular}}
\end{table}

\begin{table}[t]
\tbl{Predicted decay properties of scalar mesons.
\label{tab:prediction}} {\begin{tabular}{|c|c|c|} \hline
& \raisebox{0pt}[10pt][6pt] {Experiment} & fit  \\
\hline
 \raisebox{0pt}[10pt][6pt]{$\frac{\Gamma(f_{0}(1370)\rightarrow
K\bar{K})}{\Gamma(f_{0}(1370)\rightarrow\pi\pi )}$} &  & 0.79 \\
\hline
\raisebox{0pt}[10pt][6pt]{$\frac{\Gamma(f_{0}(1370)\rightarrow
\eta\eta)}{\Gamma(f_{0}(1370)\rightarrow K\bar{K})}$} & $0.35\pm0.30$  & 0.12\\
\hline
\raisebox{0pt}[10pt][6pt]{$\frac{\Gamma(f_2(1270)\rightarrow
\eta\eta)} {\Gamma(f_{2}(1270)\rightarrow \pi\pi)}$} &
$0.003\pm0.001$  & 0.005 \\
\hline \raisebox{0pt}[10pt][6pt]{$\frac{\Gamma(J/\psi\rightarrow
\omega f_0(1710))} {\Gamma(J/\psi\rightarrow \phi f_0(1710))}$} &
$6.6\pm2.7$  & 4.1 \\
 \hline
\raisebox{0pt}[10pt][6pt]{$\frac{\Gamma(J/\psi\rightarrow \omega
f_0(1500))} {\Gamma(J/\psi\rightarrow \phi f_0(1500))}$} &
$$  & 0.47 \\
\hline \raisebox{0pt}[10pt][6pt]{$\frac{\Gamma(J/\psi\rightarrow
\omega f_0(1370))} {\Gamma(J/\psi\rightarrow \phi f_0(1370))}$} &
$$ &  2.56 \\ \hline
\raisebox{0pt}[10pt][6pt]{$\Gamma_{f_0(1710)\to PP}$} & $<137\pm8$
 & 133 \\ \hline
\raisebox{0pt}[10pt][6pt]{$\Gamma_{f_0(1370)\to PP}$} &  & 146 \\
\hline
\end{tabular}}
\end{table}

Several remarks are in order. (i) Although $f_0(1500)$ has a large
$s\bar s$ content, the ratio of $K\bar K/\pi\pi$ is small due to
the destructive interference between the $n\bar n$ and $s\bar s$
components for $K\bar K$ production. (ii) In the absence of chiral
suppression in $G\to PP$ decay, the $f_0(1710)$ width is predicted
to be less than 1 MeV and hence is ruled out by experiment. This
is a strong indication in favor of chiral suppression of
$G\to\pi\pi$ relative to $G\to K\bar K$. (iii) Because the $n\bar
n$ content is more copious than $s\bar s$ in $f_0(1710)$, it is
natural that $J/\psi\to\omega f_0(1710)$ has a rate larger than
$J/\psi\to \phi f_0(1710)$. (iv) If $f_0(1710)$ is composed mainly
of the scalar glueball, it should be the most prominent scalar
produced in radiative $J/\psi$ decay. Hence, it is expected that
$\Gamma(J/\psi\to \gamma f_0(1710))\gg \Gamma(J/\psi\to \gamma
f_0(1500))$, a relation borne out by experiment. (v) It has been
argued that the non-observation of $f_0(1710)$ in $p\bar p$
annihilation at Crystal Barrel \cite{Amsler02} implies an $s\bar
s$ structure for the $f_0(1710)$. However,  both $f_0(1710)$ and
$f_0(1500)$ are observed in a recent study of the reaction $p\bar
p\to\eta\eta\pi^0$ at Fermlab \cite{Uman}. (vi) The $2\gamma$
coupling to scalar quarkonia has been studied in detail in
\cite{Farrar}. In our mixing model, the relative $2\gamma$
coupling strength is $f_0(1370):f_0(1500):f_0(1710)=9.3:1.0:1.5$.
Hence $f_0(1500)$ has the smallest $2\gamma$ coupling of the three
states.

\section*{Acknowledgments}
I'm grateful to Keh-Fei Liu and Chun-Khiang Chua for collaboration
on this interesting topic.

\balance

\appendix


\begin{thebibliography}{9}



\newcommand{\bib}{\bibitem}
\def\pr{{Phys. Rev.}~}
\def\prl{{Phys. Rev. Lett.}~}
\def\pl{{Phys. Lett.}~}
\def\np{{Nucl. Phys.}~}
\def\zp{{Z. Phys.}~}


\bib{Close1} C. Amsler and F.E. Close, \pl B {\bf 353}, 385
(1995); \pr D {\bf 53}, 295 (1996); F.E. Close and A. Kirk, \pl B
{\bf 483}, 345 (2000).

\bib{ams95} C. Amsler {\it et al.}, \pl B {\bf 342}, 433 (1995);
\pl B {\bf 340}, 259 (1994).

\bib{bsh93} G. Bali {\it et al.} (UKQCD), \pl B {\bf 309}, 378
(1993); C. Michael and M. Teper, Nucl. Phys. B {\bf 314}, 347
(1989).


\bib{Close2} F.E. Close and Q. Zhao, \pr D {\bf 71}, 094022 (2005).

\bib{He} X.G. He, X.Q. Li, X. Liu, and X.Q. Zeng, \pr D {\bf 73},
051502 (2006); {\it ibid.} D {\bf 73}, 114026 (2006).

\bib{Mathur} N. Mathur {\it et al.,} hep-ph/0607110.


\bib{Chen} Y. Chen {\it et al.,} \pr D {\bf 73}, 014516
(2006).

\bib{MP} C. Morningstar and M. Peardon, \pr D {\bf 56}, 3043
(1997); \pr D {\bf 60}, 034509 (1999).

\bib{CCL} H.Y. Cheng, C.K. Chua, and K.F. Liu, hep-ph/0607206.

\bib{PDG} Particle Data Group, Y.M. Yao {\it et al.,} J. Phys. G
{\bf 33}, 1 (2006).

\bib{Chanowitz} M.S. Chanowitz, \prl {\bf 95}, 172001 (2005).

\bib{Chao} K.T. Chao, X.G. He, and J.P. Ma, hep-ph/0512327.

\bib{Sexton} J. Sexton, A. Vaccarino, and D. Weingarten, \prl {\bf
75}, 4563 (1995).

\bib{Amsler02} C. Amsler {\it et al.,} Eur. Phys. J.  C {\bf 23},
29 (2002).

\bib{Uman} Uman {\it et al.,} Phys. Rev. D {\bf 73}, 052009
(2006).

\bib{Farrar} F.E. Close, G.R. Farrar, and Z. Li, Phys. Rev. D {\bf
55}, 5749 (1997).

\end{thebibliography}
\end{document}